# Heat-Assisted Multiferroic Solid-State Memory

S. Lepadatu[*,1], M.M. Vopson[2]

[1]*Jeremiah Horrocks Institute for Mathematics, Physics and Astronomy, University of Central Lancashire, Preston PR1 2HE, U.K.*

[2]*University of Portsmouth, SEES, Faculty of Science, Portsmouth PO1 3QL, U.K.*

Abstract

A heat-assisted multiferroic solid-state memory design is proposed and analysed, based on a PbNbZrSnTiO$_3$ antiferroelectric substrate and Ni$_{81}$Fe$_{19}$ magnetic free layer. Information is stored as magnetisation direction in the free layer of a magnetic tunnel junction element. The bit writing process is contactless and relies on triggering thermally activated magnetisation switching of the free layer towards a strain-induced anisotropy easy axis. A stress is generated using the antiferroelectric substrate by voltage-induced antiferroelectric to ferroelectric phase change, and this is transmitted to the magnetic free layer by strain-mediated coupling. The thermally activated strain-induced magnetisation switching is analysed here using a three-dimensional, temperature-dependent magnetisation dynamics model, based on simultaneous evaluation of the stochastic Landau Lifshitz Bloch equation and heat flow equation, together with stochastic thermal fields and magnetoelastic contributions. The magnetisation switching probability is calculated as a function of stress magnitude and maximum heat pulse temperature. An operating region is identified, where magnetisation switching always occurs, with stress values ranging from 80 to 180 MPa, and maximum temperatures normalised to the Curie temperature ranging from 0.65 to 0.99.

[*] SLepadatu@uclan.ac.uk

# 1. Introduction

Non-volatile memories for primary storage are potential candidates for a universal memory, promising both long-term storage and reliability, as well as speeds comparable to volatile memory such as dynamic random access memory (RAM). Currently the most common types of non-volatile RAM include flash memory and ferroelectric RAM [1]. Whilst these are commercially available, a number of problems prevent their use as a universal memory. Flash memory is relatively slow and unreliable due to limited number of write cycles, whilst ferroelectric RAM suffers from low bit densities. Other approaches are based on the use of magnetic materials. The most widely researched magnetic RAM is the spin transfer torque magnetic RAM (STT-MRAM) [2-4], based on switching the magnetisation direction of a free magnetic layer in a magnetic tunnel junction (MTJ) using spin-polarised currents. Whilst this is also commercially available, offering lower power consumption, faster speeds and comparable bit densities to dynamic RAM, the high manufacturing cost required to achieve large bit densities currently prevents it from being widely adopted. This stems in part from the complex multi-layered tunnel junctions in STT-MRAM. Other approaches under research include heat assisted MRAM [5], three-terminal domain wall MRAM [6] and racetrack memory [7,8]. The latter promises greatly increased areal bit densities due to a three-dimensional design allowing multiple bits to be stored per chip area.

Here a heat-assisted multiferroic memory (HAMM) device is proposed and analysed, based on a magnetoelectrical multi-layered design. Magnetisation switching at room temperature through strain-mediated coupling in multi-layered magnetoelectrical structures has been demonstrated in previous studies [9-14]. Other electric-field control methods of switching magnetisation in multiferroic structures have also been demonstrated, including electric-field control of spin polarisation [15,16], antiferromagnetic order [17] and interfacial perpendicular anisotropy in MTJs [18]. Combining both electric field control and spin polarised currents to switch the magnetisation in an MTJ has also been proposed [19]. In the HAMM array design introduced here bits are stored in MTJ elements as with MRAM, however the writing process uses a low power contactless method, based on triggering thermally activated magnetisation switching towards a strain-induced anisotropy easy axis. This avoids the difficulties encountered with STT-MRAM due to the high tunnel current densities required to induce magnetisation switching, allowing for the simplest possible MTJ stacks to be used.

## 2. Heat-Assisted Multiferroic Memory

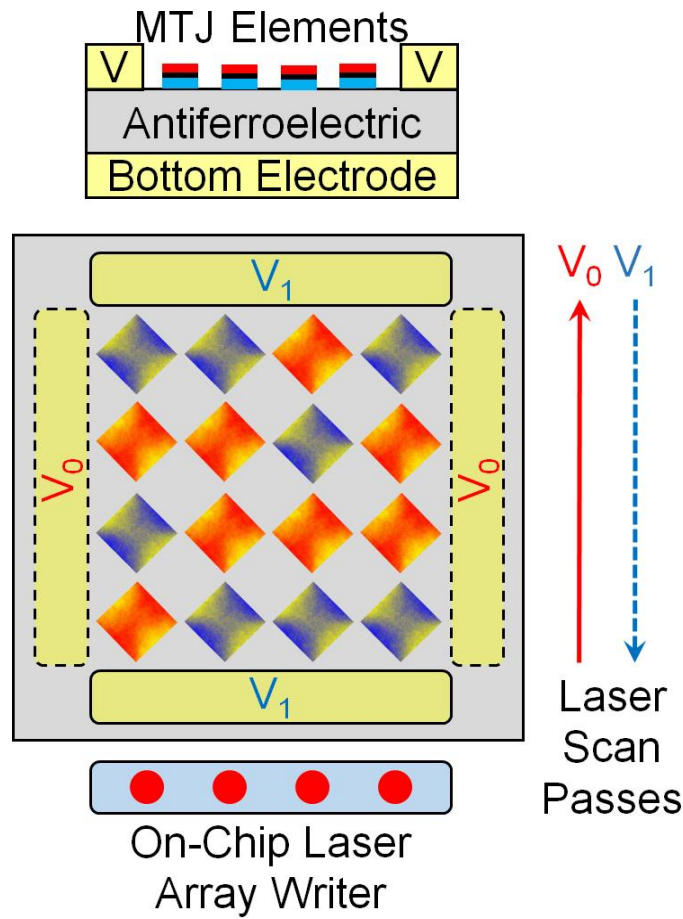

**Figure 1. HAMM array.** Information is stored in a patterned array of MTJ elements. Information is written using a low-power on-chip laser source and a minimal number of electrical contacts on the antiferroelectric substrate, by heat-assisted stress-induced magnetisation switching. Two laser scan passes are used to write a block of information, in the first pass a voltage on the $V_0$ contacts generate stress to write bits "0", whilst in the second pass bits "1" are written using the $V_1$ contacts. Heat pulses are delivered to the elements by the laser source as required during the scan passes.

The HAMM array is shown in Figure 1. The MTJ element is square shaped with bits "0" and "1" encoded as different magnetisation orientations along the two diagonals. These magnetic configurations are shown in the insets to Figure 3b-c, and also shown using color coding in the array of Figure 1. The writing process does not use direct electrical contacts to the individual elements, instead relying on a limited number of voltage pads placed on the antiferroelectric substrate as shown in Figure 1. These voltage pads take an industry standard 5 V input and through the antiferroelectric substrate a directional in-plane stress is generated over a relatively large area. A similar effect is produced by a ferroelectric substrate, but ferroelectrics display non-zero remanent polarization and strain, while anti-ferroelectrics have zero polarization and zero strain in relaxed state [20]. This condition is essential especially when the functionality of the memory cell is based on the strain mediated coupling effect, so the possibility of self-erasure or strain-induced reversal in the relaxed state is eliminated. The voltage is applied between the top and bottom electrodes, as shown in Figure 1, and the substrate thickness is chosen such that the resulting electric field strength is sufficient to induce an antiferroelectric to ferroelectric phase transition [21]. We have analysed a suitable antiferroelectric sample, PbNbZrSnTiO$_3$, with the polarisation and strain loops shown in Figure 2. Here the antiferroelectric to ferroelectric phase transition occurs above 30 kV/cm, thus for a fixed potential of 5 V a substrate thickness of ~1.5 μm is required.

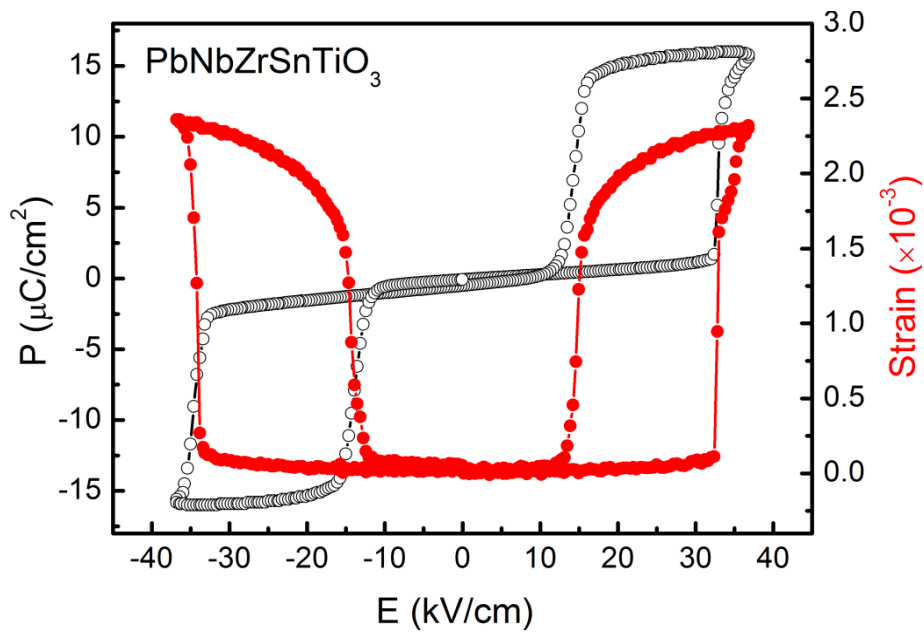

**Figure 2. Antiferroelectric substrate characterization.** Polarisation (open discs) and out-of-plane strain (closed discs) loops as a function of applied electric field are shown for an antiferroelectric PbNbZrSnTiO$_3$ substrate of 500 μm thickness. Antiferroelectric to ferroelectric phase transition occurs above 30 kV/cm.

As shown in a previous study, the electric contact geometry of Figure 1 generates an in-plane strain between the top electrode pair [22]. Two in-plane stress directions are defined using the two sets of top electrodes shown in Figure 1. Through strain-mediated coupling the stress is transmitted to a large number of MTJ elements, but crucially it is not strong enough to change the magnetisation state of the memory elements on its own. The bits are individually addressed by combining the stress input with a heat pulse delivered using a scanned pulsed laser beam. As the temperature of the magnetic layer increases, the equilibrium magnetisation length decreases, tending towards zero at the Curie temperature $T_C$. This reduces the energy barrier which must be overcome in order to switch the magnetisation configuration. In the first laser scan pass voltage $V_0$ is activated, generating an in-plane horizontal stress. In order to write bits "0" over the given array block, the scanned laser beam is turned on only over the memory elements where bits "0" must be stored. In the returning laser scan pass voltage $V_1$ is activated, now allowing all bits "1" to be written. The reading process can be done using electrical contacts to read the resistance state of the MTJ elements. Alternatively the reading process can also be contactless, using either the optical reading method demonstrated previously [23] or by using a tunnelling magneto-resistance [24] read-head element – in this case a single magnetic layer can be used instead of the MTJ multi-layer.

The advantages over other non-volatile memories, in particular STT-MRAM, include low power required for writing data, minimal number of electrical contacts and simplicity of design. There are clearly a number of engineering challenges as indicated below, although the focus here is on understanding the physical processes and their feasibility for the proposed device. Including the on-chip laser writer is an engineering challenge, although significant progress has been made in the related HAMR technology [25]. An alternative all-optical switching of MTJs using infrared laser pulses has also been proposed, relying on ferrimagnetic Gd(Fe,Co) as the free layer [26]. In order to speed-up the writing process it is desirable for the laser array writer to have multiple and independently controllable output beams. These must also be focused on the array surface and have a scanning capability as indicated in Figure 1. These requirements could be satisfied using a MEMS-based design, allowing tapping and control of multiple output laser beams from a solid-state laser, as well as scanning using built-in deformable mirrors. In order to increase the data density the magnetic elements can be reduced in size. The limitation here is set by the focused laser beam diameter, which is required to address individual elements, and this is typically around several hundred nanometres. This can be improved by careful device engineering. Since the focused laser beam has a Gaussian profile, the laser fluence is not uniform, but instead reaches a maximum at the centre. This should allow magnetisation switching of a single central MTJ element even though the laser beam diameter is larger and thus covers multiple MTJ elements – in this design the temperature reached at the outer MTJ elements is not sufficient to switch the magnetisation. Another possibility is to use a near-field laser configuration, allowing addressing of much smaller MTJ elements.

## 3. Temperature-Dependent Magnetisation Switching Modelling

In order to investigate the operation of a HAMM element, the magnetisation switching processes are investigated using a three-dimensional coupled micromagnetics model based on the stochastic Landau Lifshitz Bloch (sLLB) equation and heat flow solver as described previously [27]. The heat flow equation is given below, where $C$ is the specific heat capacity, $\rho$ is the mass density, and $K$ is the thermal conductivity. The magnetic free layer is a square of side 320 nm and 5 nm thickness with the material set as magnetostrictive Ni-rich permalloy ($Ni_{81}Fe_{19}$) and for simplicity only the free layer is simulated, placed directly on the substrate. Parameters for magnetic and thermal properties are given in Ref. [27].

$$C\rho \frac{\partial T(\mathbf{r},t)}{\partial t} = \nabla.K\nabla T(\mathbf{r},t) + Q(\mathbf{r},t) \tag{1}$$

Here $Q$ is the source term (W/m$^3$), which was set to a constant value of $1.6\times10^{18}$ W/m$^3$. For the dimensions given and a laser pulse duration of 3 ns, this can be achieved using a laser fluence of 2.4 mJ/cm$^2$, requiring a low powered laser beam of ~0.8 mW. The simulated heating and cooling cycle is shown in Figure 3a, plotting the temperature normalised to the Curie temperature for permalloy of $T_C$ = 870 K [28]. A snapshot of the temperature distribution is also shown in the inset to Figure 3a, where blue represents the lowest temperature and red the highest – the temperature is lowest at the sides of the HAMM element since there the heat loss rate to the substrate and air is highest.

The LLB equation (without the stochastic terms) is given below, where $\tilde{\alpha}_\perp = \alpha_\perp / m$ and $\tilde{\alpha}_\parallel = \alpha_\parallel / m$, with $m$ being the temperature-dependent magnetization length |**M**| normalized to its zero temperature value $M_S^0$. The transverse and longitudinal damping terms are related to the zero-temperature Gilbert damping constant $\alpha$ by $\alpha_\perp = \alpha(1 - T/3T_C)$ and $\alpha_\parallel = 2\alpha T/3T_C$.

$$\frac{\partial \mathbf{M}}{\partial t} = -\gamma \mathbf{M} \times \mathbf{H} + \frac{\tilde{\alpha}_\perp}{|\mathbf{M}|} \mathbf{M} \times \frac{\partial \mathbf{M}}{\partial t} + \gamma \frac{\tilde{\alpha}_\parallel}{|\mathbf{M}|} (\mathbf{M}.\mathbf{H})\mathbf{M} \tag{2}$$

Here $\gamma = \mu_0|\gamma_e|$, where $\gamma_e = -ge/2m_e$ is the electron gyromagnetic ratio, noting $\gamma = 2.213\times10^5$ m/As.

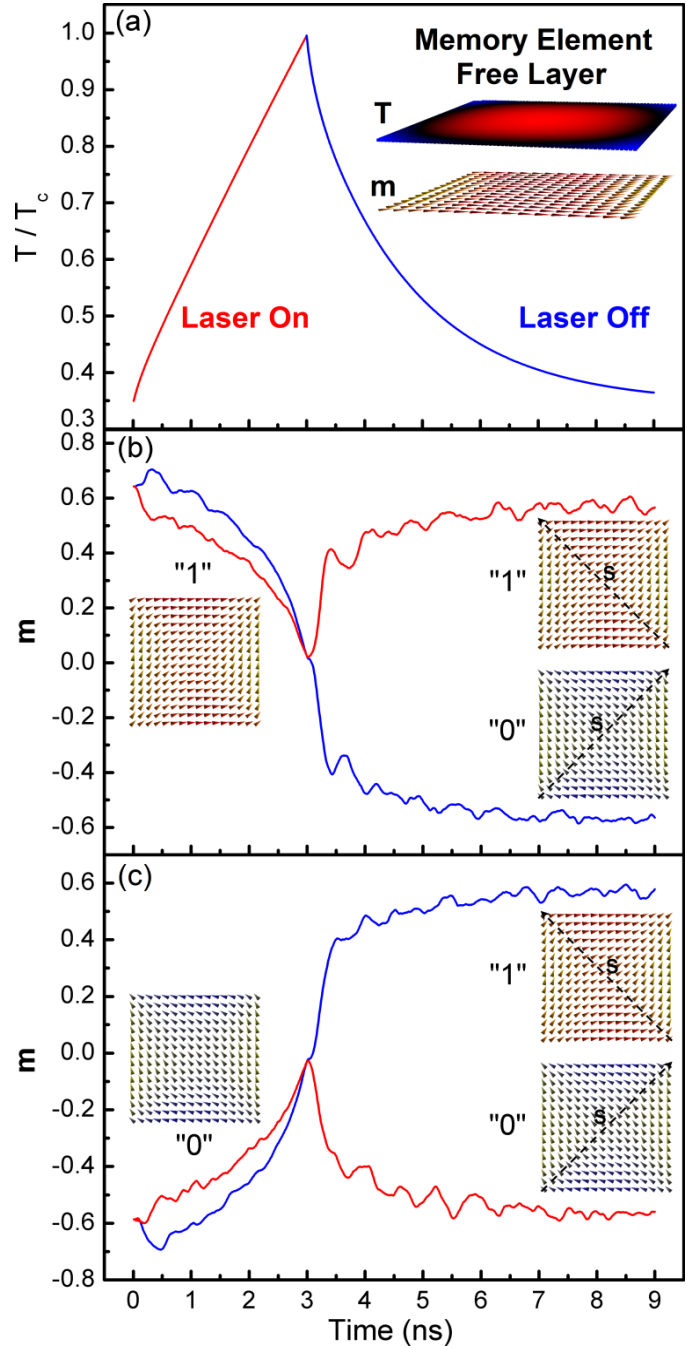

**Figure 3. Heat-assisted stress-induced switching.** (a) Heating and cooling of the MTJ free layer during and after a laser pulse. The inset shows the temperature profile in the free layer. (b),(c) Heat-assisted magnetisation switching starting from states "1" and "0" respectively for the two different stress directions, showing the magnetisation along the horizontal direction normalised to its zero-temperature saturation value as a function of time during the heating / cooling cycle.

The effective field **H** contains a number of contributions: demagnetizing field, direct exchange interaction field, external field, magnetoelastic field, as well as a longitudinal relaxation field ($T < T_C$) [29]:

$$\mathbf{H} = \mathbf{H}_{demag} + \mathbf{H}_{exch} + \mathbf{H}_{ext} + \mathbf{H}_{me} + \left(1 - \frac{m^2}{m_e^2}\right)\frac{\mathbf{M}}{\chi_\parallel} \qquad (3)$$

Here $m_e$ is the temperature-dependent equilibrium magnetization given by [29] $m_e(T) = B[m_e 3T_C/T + \mu\mu_0 H_{ext}/k_B T]$, where $\mu$ is the atomic magnetic moment (for $Ni_{80}Fe_{20}$ $\mu \cong \mu_B$ [28]), $k_B$ is the Boltzmann constant and B is the Langevin function, $B(x) = L(x) = \coth(x) - 1/x$. The longitudinal susceptibility, $\chi_\parallel$, is given by $\chi_\parallel(T) = (\partial M_e(T)/\partial H_{ext})|_{H_{ext}=0}$, where $M_e = m_e M_S^0$, thus we obtain $\chi_\parallel(T) = (\mu\mu_0 M_S^0/k_B T) B'(x) / (1 - B'(x)3Tc/T)$, where $x = m_e 3Tc/T$, and B′ is the differential of the Langevin function. The exchange field is given by $\mathbf{H}_{exch} = (2A(T)/\mu_0 M_e^2) \nabla^2 \mathbf{M}$, where $A(T) = A_0 m_e^2(T)$ [30], $A_0$ being the zero-temperature value of the exchange stiffness, $A_0 = 1.3 \times 10^{-11}$ J/m for permalloy.

The magnetoelastic field is derived from the magnetoelastic energy density [31], given in equation (4), using the expression $\mathbf{H}_{me} = -1/\mu_0 M_e \, \partial\varepsilon/\partial\mathbf{m}$.

$$\varepsilon_{me} = -\frac{3}{2}\lambda_{100}\sigma\left[\alpha_1^2\gamma_1^2 + \alpha_2^2\gamma_2^2 + \alpha_2^2\gamma_2^2\right] - 3\lambda_{111}\sigma\left[\alpha_1\alpha_2\gamma_1\gamma_2 + \alpha_2\alpha_3\gamma_2\gamma_3 + \alpha_3\alpha_1\gamma_3\gamma_1\right] \qquad (4)$$

Here $\lambda_{100}$ and $\lambda_{111}$ are the magnetostriction coefficients along the crystallographic axes, $\sigma$ is the stress generated by the antiferroelectric substrate and transmitted through to the magnetic layer by strain-mediated coupling, $\mathbf{m} = (\alpha_1, \alpha_2, \alpha_3)$ and $(\gamma_1, \gamma_2, \gamma_3)$ are the direction cosines of the magnetisation and stress respectively. Here for simplicity the magnetostriction is assumed to be isotropic with $\lambda = \lambda_{100} = \lambda_{111} = -10^{-5}$ [32] and any temperature dependence is not taken into consideration. A uniform compressive stress is used here, initially fixed to $\sigma = 100$ MPa. In-plane stress values of this order can easily be achieved in the geometry of Figure 1 using the $PbNbZrSnTiO_3$ antiferroelectric substrate. The stress tensor is obtained from the product of the elastic constant, $c^E$, and strain tensors. For the out-of-plane strain measured in Figure 2, a simple estimation using $c_{13} \cong 85$ GPa [22] results in a maximum achievable in-plane stress of ~200 MPa.

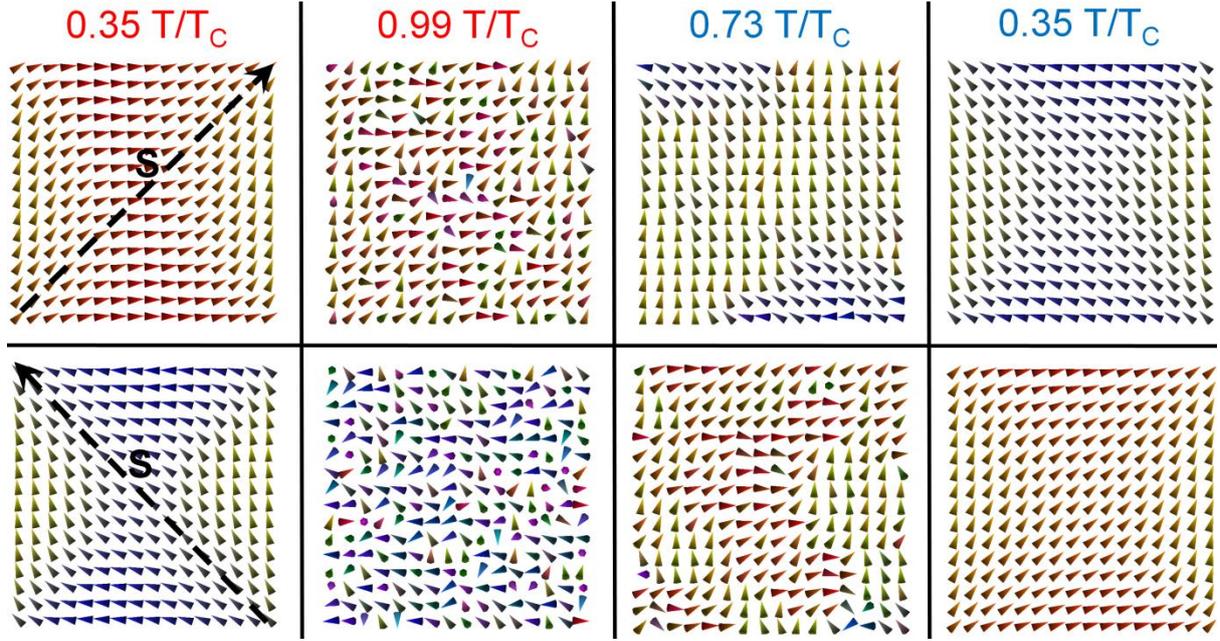

**Figure 4. Magnetisation switching simulations.** Snapshots of the magnetisation configuration are shown at different temperatures, illustrating the switching process. The top row starts from state "1" with a stress applied to induce switching to state "0". The bottom row starts from state "0" and switches to state "1".

The magnetisation switching process is shown in Figure 3b-c starting from the "1" and "0" states respectively. Depending on the applied stress direction the end state is either "0" or "1" respectively as indicated in the figure. The applied stress induces an easy axis along the opposite diagonal, however at lower temperatures this is not strong enough to result in magnetisation switching. As the Curie temperature is approached the average magnetisation length approaches zero. This reduces the effective energy barrier which must be overcome in order for the magnetisation to switch under the effect of the induced anisotropy due to the magnetoelastic coupling. To capture this process it is important to include the effect of lattice vibrations on the magnetisation due to the non-zero temperature. This is achieved using the stochastic LLB equation (sLLB), where a thermal field, $\mathbf{H}_{th}$, is added to the transverse damping torque effective field in the explicit form of the sLLB equation, and a thermal torque, $\mathbf{\eta}_{th}$, is added to the sLLB equation [33]. The thermal field and torque are given in Eq. (5), where their spatial and cross-correlations are zero, $V$ is the volume of the computational cellsize which was set to 5 nm$^3$, $\Delta t$ is the fixed time-step used in the sLLB evaluation (the Milstein scheme was used here with $\Delta t = 0.1$ ps [34]) and $\mathbf{r}_H$, $\mathbf{r}_\eta$ are random unit vectors.

$$\mathbf{H}_{th} = \frac{1}{\alpha_\perp}\sqrt{\frac{2k_B T(\alpha_\perp - \alpha_\parallel)}{\gamma\mu_0 M_S^0 V \Delta t}}\mathbf{r}_H \quad (\text{A/m}) \tag{5}$$

$$\boldsymbol{\eta}_{th} = \sqrt{\frac{2k_B T \alpha_\parallel \gamma M_S^0}{\mu_0 V \Delta t}}\mathbf{r}_\eta \quad (\text{A/ms})$$

For magnetic nanoparticles the switching probability can be described using an Arrhenius law based on the Néel-Brown thermal activation model [35,36]. Here the switching process tends to be dominated by reverse domain nucleation at the corners. This is illustrated in Figure 4, where snapshots of the magnetisation configuration during switching events are shown at different temperatures in the heating-cooling cycle. Close to $T_C$, due to the strong effect of lattice vibrations on the magnetisation, the configuration is almost random, although a preferential alignment along the starting magnetisation configuration is still maintained. As the sample cools reverse domains are nucleated along the induced anisotropy axis. The reversed domains quickly grow in size, finally reaching the reversed magnetisation configuration along the opposite diagonal.

To investigate the switching process further, the switching probability is calculated as a function of stress magnitude and maximum temperature during the heating-cooling cycle. Keeping the same laser fluence, this is adjusted by controlling the duration of the laser pulse. For each combination of stress and maximum temperature the switching probability is calculated out of 5 heating-cooling cycles. The resultant switching probability is shown in Figure 5. As expected at stronger stress values and temperatures magnetisation switching always occurs, defining the possible operating region for the HAMM array. A linear boundary delineates this region up to ~200 MPa stress where magnetisation switching can occur even at room temperature given enough time. If the temperature reaches values very close to $T_C$ the starting magnetisation configuration is completely lost and the magnetisation recovery process becomes more complex. In particular vortex structures tend to be nucleated, preventing the magnetisation from aligning along one of the diagonals. This results in a switching probability which is only weakly influenced by the applied stress magnitude as seen in Figure 5. Note the antiferroelectric transition temperature of the substrate is typically lower than the Curie temperature of $Ni_{81}Fe_{19}$, thus the higher operating temperatures may need to be avoided; it should be noted however the temperature of the substrate is lower than that of the top magnetic layer, due to radiative heat loss to air and steep temperature gradient along the substrate thickness. This can be further minimised by insertion of a thermally insulating spacer layer between the MTJ and antiferroelectric substrate.

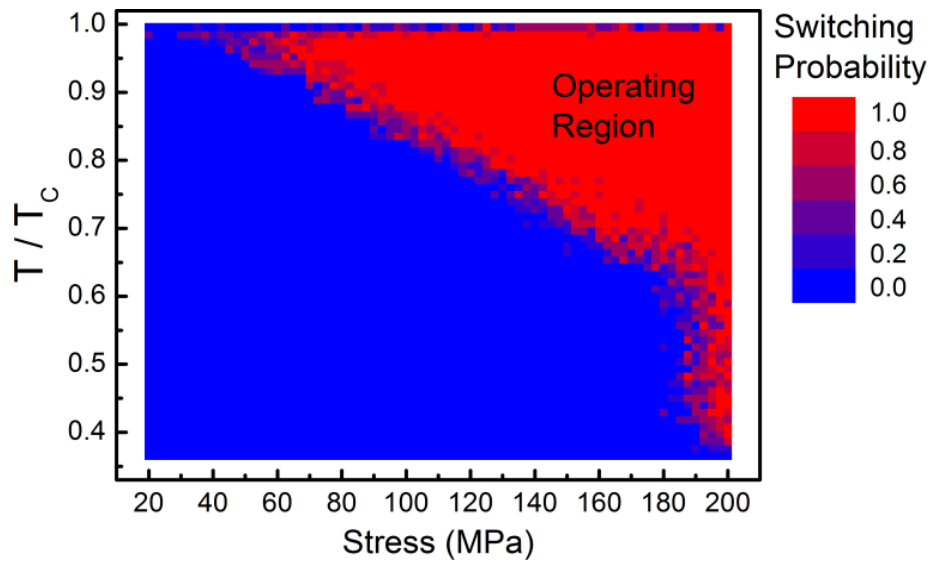

**Figure 5. Switching probability as a function of temperature and stress magnitude.** The probability of switching from state "1" to state "0" was computed as a function of maximum heat pulse temperature (varied by changing the laser pulse duration) from room temperature up to $T_C$, and as a function of stress magnitude. The operating region where switching always occurs is marked.

## 4. Conclusions

A bilayer magnetoelectric memory device has been investigated. The device consists of an antiferroelectric substrate, used to generate stresses using a minimal number of voltage pads, with information stored as the in-plane magnetisation direction in a magnetic free layer placed on the antiferroelectric substrate. Heat pulses are generated using a low powered laser, and these are used to trigger thermally activated switching of the magnetisation towards a strain-induced anisotropy easy axis. A three-dimensional micromagnetics solver based on the stochastic LLB equation and coupled to a heat flow solver has been used to investigate the magnetisation switching processes in these devices. The switching probability depends on both the applied stress magnitude and maximum temperature reached during the heat pulse, defining an operating region between 80 – 180 MPa and normalised temperatures ranging from 0.65 to 0.99. These results demonstrate the physical processes behind the proposed memory device. This simple architecture retains the advantages of STT-MRAM, namely non-volatility, fast bit reading and writing, reliability and low power usage, whilst avoiding problems related to the high tunnel current densities required for switching the magnetisation in MTJ elements. Whilst the proposed design allows for a simple architecture of the magnetoelectric layers the most important difficulty that must be overcome is increasing the areal bit density. Below a certain element size thermal stability becomes a concern and out-of-plane magnetisation devices may need to be investigated. Before this limit is reached the limitation rests with the minimum element size that can be addressed using a laser spot in order to deliver a heat pulse. The possibility of tuning the operating region by taking into account the non-uniform laser fluence has been discussed. Other possibilities include use of a near-field laser design based on a MEMS architecture, or delivering localised heat pulses using an alternative method.